\documentclass[aps,12pt,superscriptaddress,preprint]{revtex4}
 
\usepackage[dvips]{graphicx}
\usepackage{amsmath,amssymb,amsthm,bm}
 
\newcommand{\N}{\mathbb{N}}
\newcommand{\R}{\mathbb{R}}
\newcommand{\Z}{\mathbb{Z}}

\begin{document}

\preprint{IUHET-517}

\vspace*{0.75in}

\title{The Aharonov-Bohm effect for a knotted magnetic solenoid}

\author{Roman V. Buniy}
\email{roman.buniy@gmail.com}
\affiliation{Physics Department, Indiana University, Bloomington, IN 47405}

\author{Thomas W. Kephart}
\email{tom.kephart@gmail.com}
\affiliation{Department of Physics and Astronomy, Vanderbilt
University, Nashville, TN 37235} 

\date{August 13, 2008}

\begin{abstract}
 We show that the linking of a semiclassical path of a charged
 particle with a knotted magnetic solenoid results in the
 Aharonov-Bohm effect. The phase shift in the wave function is
 proportional to the flux intersecting a certain connected and
 orientable surface bounded by the knot (a Seifert surface of the
 knot).
\end{abstract}


\maketitle

\section{Introduction}

The magnetic Aharonov-Bohm effect~\cite{Aharonov:1959fk} results when
a charged particle travels around a closed path in a region of
vanishing magnetic field but nonvanishing vector potential. The
magnetic flux is confined to a region where the particle is excluded,
but the wave function of the particle is nonetheless affected by the
vector potential and an interference pattern occurs at a detection
screen.

\begin{figure}[ht]
  \includegraphics[width=7.5cm]{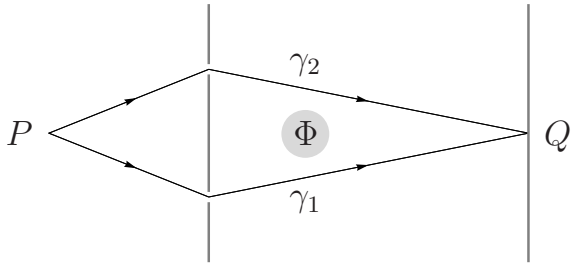}
  \caption{\label{figure-ab-apparatus} A plane projection of the
    standard magnetic Aharonov-Bohm effect apparatus. The electrons
    travel along semiclassical paths $\gamma_1$ and $\gamma_2$ from
    the source $P$ to the detection screen $Q$. The electrons do not
    penetrate the interior of the solenoid which carries the flux
    $\Phi$.}
\end{figure}

Let $C$ be the center line of a magnetic solenoid with the enclosed
flux $\Phi$ and the vector potential $\bm{A}$. For an electron moving
along a semiclassical path $\gamma$, the wave function is
$\psi(\bm{A})=\psi(0)\exp{(i\xi\int_\gamma\bm{A}\cdot d\bm{x})}$,
where $\xi=e(\hbar c)^{-1}$. For a standard magnetic Aharonov-Bohm
effect apparatus in Fig.~\ref{figure-ab-apparatus}, the total
amplitude for paths $\gamma_1$ and $\gamma_2$ is $\psi({\bm A})
=e^{i\theta}(\psi_1(0)+e^{-i\phi}\psi_2(0))$. The quantity $\theta$ is
an overall irrelevant phase and the quantity $ \phi=\xi\int_\gamma{\bm
A}\cdot d\bm{x}$, where $\gamma=\gamma_1\gamma_2^{-1}$, measures the
relative phase shift between paths $\gamma_1$ and $\gamma_2$. Applying
the Stokes theorem, we find $\phi=\xi\Phi$.

There is a certain duality in this picture. Namely, if we take
$\gamma$ to be the center line of a solenoid with the magnetic flux
$\Phi_*$ and $C$ a path of an electron, then the phase shift is
$\phi_*=\xi\Phi_*$. (We close the path $C$ far away from the
apparatus, e.g., at infinity.) The reason for this is the fact that
the ratio of the phase to the flux is proportional to the gaussian
linking of the curves $\gamma$ and $C$, which is symmetric with
respect to the curves. We will later use this duality in the more
complicated case of knotted curves.

The phase $\phi$ is of course gauge invariant. For explicit
computations, however, it is convenient to choose a singular
gauge~\cite{Buniy:2006tq,Buniy:2006tr,Buniy:2006ts} in which
$\bm{A}=\xi\Phi\delta_S\bm{n}_S$. Here $S$ is a connected and
orientable surface for which the curve $C$ is the boundary, $\delta_S$
is the delta function with the support on $S$, and $\bm{n}_S$ is the
unit vector normal to $S$. (For an infinite solenoid, $S$ is the half
plane. For a toroidal solenoid, $S$ is a disk.) It is clear that each
time a closed path $\gamma$ intersects the surface $S$, the quantity
$\int_{\gamma}\bm{A}\cdot d\bm{x}$ increases or decreases by the
quantity $\Phi$ depending on whether the intersection of $\gamma$ and
$S$ is positive or negative. Thus, for an arbitrary path $\gamma$, the
phase is $\phi=N\xi\Phi$, where the integer $N$ is the signed number
of times $\gamma$ intersects $S$. In the dual picture, we consider a
connected and orientable surface $\sigma$ for which the curve $\gamma$
is the boundary, choose the gauge potential
$\bm{A}_*=\xi\Phi_*\delta_\sigma\bm{n}_\sigma$, and obtain
$\phi_*=N_*\xi\Phi_*$, where $N_*$ is the signed number of times $C$
intersects $\sigma$. A crucial observation is that the two
intersection numbers are equal, $N=N_*$.

For a fixed closed curve $C$, any possible path of an electron belongs
to $C'=\R^3\backslash C$, the complement of $C$ in $\R^3$. The set of
all such paths form a group $\Gamma$ under the operation of
multiplication of paths. This group is $\Gamma=\pi_1(C')$, the first
homotopy group~\cite{Hatcher} of $C'$, also called the fundamental
group of $C'$.

Topologically, $C$ bounds a disk. If we continuously deform this disk,
the signed intersection number does not change. This can be seen by
noting that during deformations of $S$, new intersection points of
$\gamma$ and $S$ appear in pairs, and the two points in each pair have
intersections of opposite signs. This means that $N$ depends only on
the topological class to which $\gamma\in\Gamma$ belongs. It follows
from the above that $\Gamma\cong\Z$. In the dual picture, we need
$\Gamma_*=\pi_1(\R^3\backslash\gamma)$, the fundamental group of the
complement of $\gamma$, and it similarly follows that
$\Gamma_*\cong\Z$. In the next section we will generalize these ideas
to orientable surfaces which are bounded by knots.

The above example has a simple topology, which resulted in abelian
groups $\Gamma$ and $\Gamma_*$. The Aharonov-Bohm analysis can be
extended to examples of more complicated topologies. One possibility
is to consider multiple magnetic solenoids, which might be unlinked or
linked with each other. We have studied this case in
Refs.~\cite{Buniy:2006tq,Buniy:2006tr}, where we showed that the phase
is proportional to the product of fluxes from different solenoids and
depends on linking numbers of higher orders. Our purpose here is to
consider a simpler case of one self-knotted closed solenoid.

\section{Knots}

\begin{figure}[ht]
  \includegraphics[width=6cm]{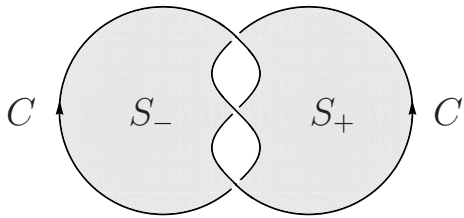}
  \caption{\label{figure-seifert-surface} The trefoil knot $C$ and its
  Seifert surface $S$. $S_+$ and $S_-$ are the two sides of the
  orientable $S$. If a semiclassical path of an electron intersects
  $S$, then the Aharonov-Bohm effect results.}
\end{figure}

We now consider the case of a closed self-knotted curve $C$. For each
such $C$, there are surfaces which are bounded by $C$. If such a
surface is connected and orientable, then it is called a Seifert
surface of $C$. A well-known theorem~\cite{Rolfsen} states that there
is a Seifert surface for every knot. (Note that some knots also bound
non-orientable surfaces; for example, there is a Mobius strip with a
$3\pi$ twist which is bounded by the trefoil knot. This is not a
Seifert surface, and we will have no use for such non-orientable
surfaces here.) In general, for a given $C$, there can be more than
one nonequivalent Seifert
surface~\cite{Seifert-surfaces,Rolfsen}. However, the signed number of
intersections of a closed curve $\gamma$ and $S$ is independent of the
choice of $S$. Hence our results will be independent of the choice of
a Seifert surface. A Seifert surface for the trefoil knot is shown in
Fig.~\ref{figure-seifert-surface}.

A generalization of the Stokes theorem for knotted closed
curves~\cite{PRB} states that $\int_\gamma \bm{A}\cdot d\bm{x }
=N\Phi$, where $N$ is the signed intersection number of $\gamma$ and
$S$, and $S$ is a Seifert surface of the knot $C$. In the dual picture
with the unknotted solenoid $\gamma$ with flux $\Phi_*$ and a knotted
semiclassical path of an electron $C$, we have $\int_C \bm{A}_*\cdot
d\bm{x } =N_*\Phi_*$, where $N_*$ is the signed intersection number of
$C$ and $\sigma$, and $\sigma$ is a Seifert surface of $\gamma$. As in
the case of unknotted closed curves, the numbers $N$ and $N_*$ are
equal since they both again represent the gaussian linking of the
curves $\gamma$ and $C$.

\begin{figure}[ht]
  \includegraphics[width=5cm]{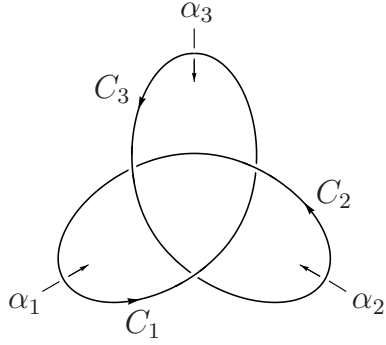}
  \caption{\label{figure-trefoil-wirtinger-presentation} A picture of
  the trefoil knot with the arcs $C_1$, $C_2$, $C_3$ and its Wirtinger
  presentation with the vectors $\alpha_1$, $\alpha_2$, $\alpha_3$.}
\end{figure}

\begin{figure}[ht]
  \includegraphics[width=10cm]{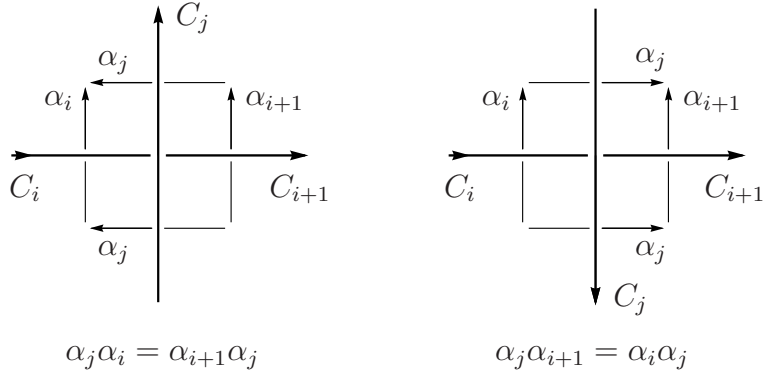}
  \caption{\label{figure-wirtinger-relations} The Wirtinger relations
  for a knot.}
\end{figure}

For the fundamental group $\Gamma$ with generators
$\alpha_1,\ldots,\alpha_n$ satisfying relations
$\beta_1=1,\ldots,\beta_n=1$, we write~\cite{Rolfsen}
\begin{align}
  \Gamma=(\alpha_1,\ldots,\alpha_n ; \beta_1,\ldots,\beta_n).
\end{align}
The generators $\{\alpha_i\}$ and relators $\{\beta_i\}$ can be found
for any knot $C$ by using the Wirtinger presentation as follows. Let
there be given a picture of the knot $C$ as a set of arcs
$C_1,\ldots,C_n$ in a plane. Each $C_i$ is assumed to be connected
with $C_{i-1}$ and $C_{i+1}$ by undercrossing arcs as in
Fig.~\ref{figure-trefoil-wirtinger-presentation}. Indices are taken
modulo $n$. We assume that the arcs are oriented in the order of
$C_1,\ldots,C_n$. For each arc $C_i$, we draw a vector $\alpha_i$
crossing $C_i$ with the fixed orientation relative to the orientation
of $C_i$. At each of the $n$ crossings, there are two possibilities as
in Fig.~\ref{figure-wirtinger-relations}. These lead to the relations
$\alpha_j\alpha_i=\alpha_{i+1}\alpha_j$ and
$\alpha_j\alpha_{i+1}=\alpha_i\alpha_j$. It can be
proved~\cite{Rolfsen} that there are no other relations for the group
$\Gamma$. This means that a relator $\beta_i$ equals either
$\alpha_j\alpha_i\alpha_j^{-1}\alpha_{i+1}^{-1}$ or
$\alpha_j\alpha_{i+1}\alpha_j^{-1}\alpha_i^{-1}$. Since there is a
product of all $\beta$s which is trivially the identity, any one of
the relations defining $\Gamma$ can be omitted.

Keeping the orientation and notation of
Fig.~\ref{figure-wirtinger-relations}, we can close the $\alpha$ lines
to form circles. Then each element $\gamma\in\Gamma$ can be written in
the form~\cite{free-groups}
\begin{align}
  \gamma=\alpha_1^{k_{1,1}}\cdots\alpha_n^{k_{n,1}} \cdots
  \alpha_1^{k_{1,l}}\cdots\alpha_n^{k_{n,l}},
\end{align}
where $l\in\N$, $k_{i,j}\in\Z$.

Since the first homotopy group $\Gamma=\pi_1(C')$ is now nonabilian
and phases in quantum mechanics are elements of the abelian group
$\mathrm{U}(1)$, we need to find $G=H_1(C')$, the first homology group
of $C'$, which is the abelianization of
$\pi_1(C')$~\cite{Buniy:2006tq,Buniy:2006tr}. We obtain elements of
$G$ by considering elements of $\Gamma$ modulo their
commutators~\cite{Hatcher}. This means that we obtain $G$ from
$\Gamma$ by replacing the non-commutative generators $\{\alpha_i\}$
and relators $\{\beta_i\}$ by commutative generators $\{a_i\}$ and
relators $\{b_i\}$, respectively,
\begin{align}
  G=(a_1,\ldots,a_n;b_1,\ldots,b_n).
\end{align}
Here $b_i$ equals either $a_ia_{i+1}^{-1}$ or $a_{i+1}a_i^{-1}$, which
means that the corresponding relations allow us to replace $a_{i+1}$
by $a_i$, or vice versa. As a result, the element $c\in G$
corresponding to $\gamma\in\Gamma$ is $c=a_h^m$, where $h$ is any
number from the set $\{1,\ldots,n\}$ and
$m=\sum_{i=1}^n\sum_{j=1}^lk_{i,j}$. Since $m\in\Z$, this implies that
$G\cong\Z$ and $a_h$ is the corresponding meridional generator.

\begin{figure}[ht]
  \includegraphics[width=6.5cm]{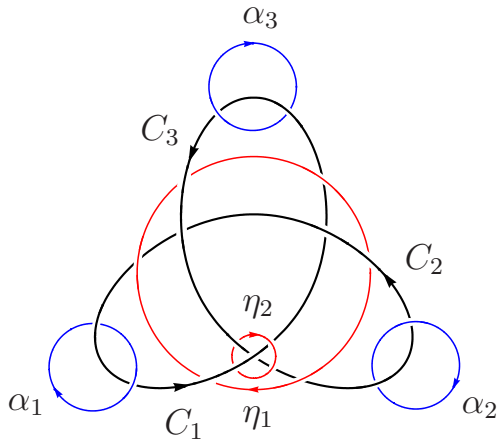}
  \caption{\label{figure-trefoil-homotopy-generators} The first
    homotopy group $\pi_1(T'_{2,3})$ is generated by the generators
    $\alpha_1$, $\alpha_2$, $\alpha_3$ which satisfy the relations
    given by Eq.~\eqref{trefoil-wirtinger-relators}. Alternatively,
    $\pi_1(T'_{2,3})$ is generated by the generators $\eta_1$,
    $\eta_2$, which satisfy the relation $\eta_1^2=\eta_2^3$.}
\end{figure}

Since $G$ is an abelian group, the phase $\phi=N\xi\Phi$ is additive
for multiplicative paths, as it must be in quantum
mechanics~\cite{path-integral}, and now $N$ is the number of times
$\gamma$ intersects the Seifert surface of the knot.

One of the simplest classes of nontrivial knots is the torus knots
$\{T_{p,q}\}_{p,q\in\Z}$. The torus knot $T_{p,q}$ wraps around the
solid torus in the longitudinal direction $p$ times and in the
meridional direction $q$ times. We require that the numbers $p$ and
$q$ are coprime and $\vert p\vert\not=1$, $\vert q\vert\not=1$, since
otherwise $C$ is unknotted. The simplest example in the family of
torus knots is the trefoil knot $T_{2,3}$; see
Fig.~\ref{figure-trefoil-wirtinger-presentation}.

\begin{figure}[ht]
  \includegraphics[width=7.5cm]{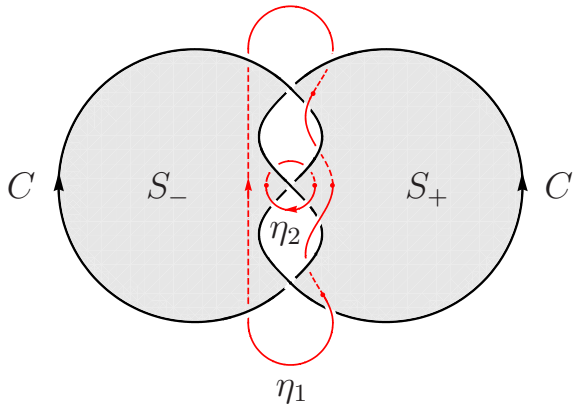}
  \caption{\label{figure-trefoil-seifert-generators} An alternative
   view of the trefoil knot, where the generators $\eta_1$ and
   $\eta_2$ of $\pi_1(T'_{2,3})$ are shown intersecting the Seifert
   surface thrice and twice, respectively. The relation
   $\eta_1^2=\eta_2^3$ implies that deforming $\eta_1^2$ into
   $\eta_2^3$ conserves the number of intersections of the path with
   the Seifert surface, which in this case is equal to six.}
\end{figure}

The simplest presentation of the fundamental group of
$T'_{p,q}=\R^3\backslash T_{p,q}$ is
\begin{align}
  \Gamma=(\eta_1,\eta_2 ; \eta_1^p\eta_2^{-q}).
\end{align}
For the trefoil knot, the Wirtinger relators are
\begin{align}
  \beta_1=\alpha_3\alpha_2\alpha_3^{-1}\alpha_1^{-1}, \quad
  \beta_2=\alpha_1\alpha_3\alpha_1^{-1}\alpha_2^{-1}, \quad
  \beta_3=\alpha_2\alpha_1\alpha_2^{-1}\alpha_3^{-1},
  \label{trefoil-wirtinger-relators}
\end{align}
where the Wirtinger generators are related to the generators of
$\pi_1(T'_{2,3})$ via $\eta_1=\alpha_1\alpha_3\alpha_2$,
$\eta_2=\alpha_2\alpha_1$; see
Fig.~\ref{figure-trefoil-homotopy-generators}. Using the $\beta_3$
Wirtinger relation to eliminate $\alpha_3$, we have
$\eta_1=\alpha_1\alpha_2\alpha_1$. It is now straightforward to check
that $\eta_1^2=\eta_2^3$~\cite{footnote}. (The reader might find it
useful to experiment with wires and strings to verify these results.)
In general, the relations in the fundamental group preserve the signed
intersection number of the closed path with the Seifert surface of the
knot. For the trefoil knot example, see
Fig.~\ref{figure-trefoil-seifert-generators}.

\begin{figure}[ht]
  \includegraphics[width=6cm]{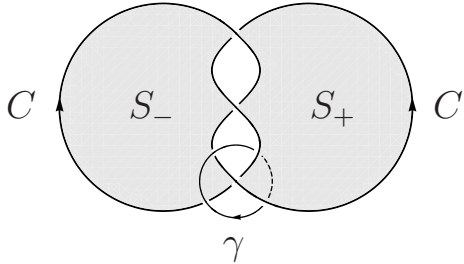}
  \caption{\label{figure-seifert-link} An example of an electron path
  $\gamma$ which does not intersect the Seifert surface $S$ but
  nevertheless links with the knot $C$.}
\end{figure}

Note that there are paths, for example, $\gamma$ in
Fig.~\ref{figure-seifert-link}, that are linked with the knot, but
that do not intersect the Seifert surface.
Hence, if the path of an electron is linked through one of the holes
in the Seifert surface, then there is no gaussian linking, and no
standard Aharonov-Bohm effect. However, these paths have higher order
linking with the knot and may result in a higher order effects where
$\phi$ is nonlinear in flux as discussed in
Refs.~\cite{Buniy:2006tq,Buniy:2006tr}. The simplest example of higher
order linking, the Borromean rings with an electron semiclassical path
linking in a specific way with two unlinked solenoids, leads to a
phase that is second order in fluxes~\cite{Buniy:2006tq}.

\section{Conclusion}

We conclude that the nontrivial phase that can be detected in the case
of a knotted solenoid is $\phi=N(e\Phi)/(\hbar c)$, where $N$ is the
number of times the semiclassical path of an electron intersects a
Seifert surface $S$ of the knot $C$, and $\Phi$ is the magnetic flux
within the knotted solenoid. This generalizes the case of a simple
toroidal solenoid, but continues to correspond to the gaussian linking
of the semiclassical particle path and the magnetic solenoid. Note
that combining the methods developed here and in
Refs.~\cite{Buniy:2006tq,Buniy:2006tr}, we in principle know the
quantum-mechanical phase for the case of multiple knotted magnetic
solenoids and a knotted path of an electron.

We will not suggest an experimental setup for detecting the
Aharonov-Bohm effect for knots, since this is better left to those
with the technical expertise who know best how to carry out such an
experiment. However, it may be useful to approach the problem from the
point of view of the Josephson effect where analogous generalized
Aharonov-Bohm experiments~\cite{BK} can be carried out.

\begin{acknowledgments}

The work of RVB was supported by DOE grant number DE-FG02-91ER40661
and that of TWK by DOE grant number DE-FG05-85ER40226.

\end{acknowledgments}


\begin{thebibliography}{100}
\bibitem{footnote} For comparison, the $\pi_1(T'_{2,3})$ generators
$\eta_1$ and $\eta_2$ of Fig. \ref{figure-trefoil-homotopy-generators}
were called $\beta$ and $\alpha$, respectively, in Ref.~\cite{PRB}.
\end{thebibliography}
\end{document}